\documentclass[preprint,5p,times,onecolumn]{elsarticle}

\usepackage{graphicx}				
\usepackage[utf8]{inputenc}				
\usepackage{axodraw4j}
\usepackage{pstricks}
\usepackage{xspace}
\usepackage{amssymb}
\usepackage{amsmath}
\def\eg{\emph{e.g.}}
\def\ie{\emph{i.e.}}

\def\etal{\emph{et~al.}}

\newcommand{\pythia}{P\protect\scalebox{0.8}{YTHIA}8\xspace}

\journal{Physics Letters B}

\begin{document}
\title{Soft modifications to jet fragmentation in high energy proton--proton collisions}
\author[adrlabel,adrlabel2]{Christian Bierlich\corref{cor1}}
\address[adrlabel]{Niels Bohr Institute, University of Copenhagen, Blegdamsvej 19, 21000 København Ø, Denmark.}
\address[adrlabel2]{Department of Astronomy and Theoretical Physics, Lund University, S{\"o}lvegatan 14A, S 223 62 Lund, Sweden.}
\cortext[cor1]{Corresponding author:\\\textit{E-mail:} christian.bierlich@thep.lu.se \\ \textit{Postal:} Department of Astronomy and Theoretical Physics, Lund University, S{\"o}lvegatan 14A, S 223 62 Lund, Sweden. \\ \textit{Preprint numbers:} LU-TP 19-05, MCnet-19-01, 1901.07447 [hep-ph].}


\begin{frontmatter}
\begin{abstract}
The discovery of collectivity in proton--proton collisions, is one of the most puzzling outcomes from the first two runs at LHC, as it points to the possibility of creation of a Quark--Gluon Plasma, earlier believed to only be created in heavy ion collisions. One key observation from heavy ion collisions is still not observed in proton--proton, namely jet-quenching. In this letter it is shown how a model capable of describing soft collective features of proton--proton collisions, also predicts modifications to jet fragmentation properties. With this starting point, several new observables suited for the present and future hunt for jet quenching in small collision systems are proposed.
  
\end{abstract}

\end{frontmatter}

\section{Introduction}%
One of the key open questions from Run 1 and Run 2 at the LHC, has been prompted by the observation
of collective features in collisions of protons, namely the observation of a 
near--side ridge \cite{Khachatryan:2010gv}, as well as strangeness enhancement with multiplicity
\cite{ALICE:2017jyt}.
Similar features are, in collisions of heavy nuclei, taken as evidence 
for the emergence of a Quark--Gluon Plasma (QGP) phase, few fm after
the collision.

The theoretical picture of collective effects in heavy ion collisions is vastly
different from the picture known from proton--proton (pp). Due to the very different geometry of 
the two system types, interactions in the final state of the collision become
dominant in heavy ion collisions, while nearly absent in pp collisions. The 
geometry of heavy ion collisions is so different from pp collision that in 
fact even highly energetic jets suffer an energy loss traversing the medium, a phenomenon
known as jet quenching.

The ATLAS experiment has recently shown that the ridge remains in
events tagged with a $Z$-boson \cite{ATLAS:2017nkt}. While maybe unsurprising by itself,
the implication of this measurement is a solid proof that \emph{some} 
collective behaviour exists in events where a high-$p_\perp$ boson is produced, possibly with
an accompanying jet. In this letter this observation is taken as a 
starting point to investigate how the same dynamics producing the ridge in $Z$-tagged
collisions, may also affect jet fragmentation. To that end, the microscopic model
for collectivity, based on interacting strings \cite{Bierlich:2017vhg,Bierlich:2016vgw,Bierlich:2014xba}
is used. The model has been shown to reproduce the near side ridge in minimum bias pp, 
and has been implemented in the \pythia event generator \cite{Sjostrand:2014zea}, allowing one to study 
its influence also on events containing a $Z$ and a hard jet.

The non-observation of jet quenching in pp and pPb collisions is, though
maybe not surprising due to the vastly different geometry, one of the most puzzling
features of small system collectivity. If collectivity in small systems is due to
final state interactions, it should be possible to also measure its effect on jets.
If, on the other hand, collectivity in small collision systems is \emph{not} due
to final state interactions, but mostly due to saturation effects in the initial state
 -- as predicted by Color Glass Condensate calculations \cite{Schenke:2016lrs} -- the non-observation
 of jet quenching will follow by construction. The continued search for jet quenching
 in small systems is therefore expected to be a highly prioritized venue for the
 upcoming high luminosity phase of LHC \cite{Citron:2018lsq}.

\begin{figure*}
	\hspace{1cm}
	\includegraphics[width=0.35\textwidth]{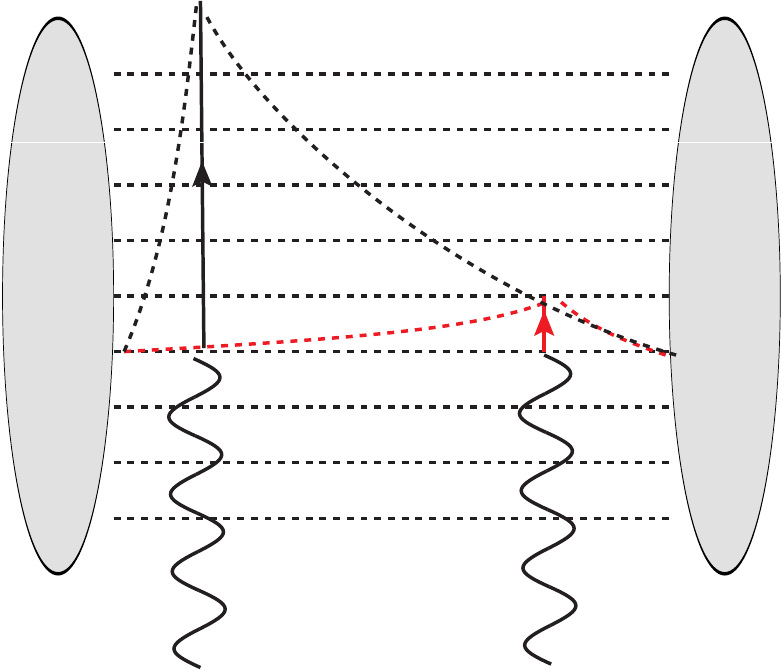}
	\hspace{2.5cm}
	\includegraphics[width=0.4\textwidth]{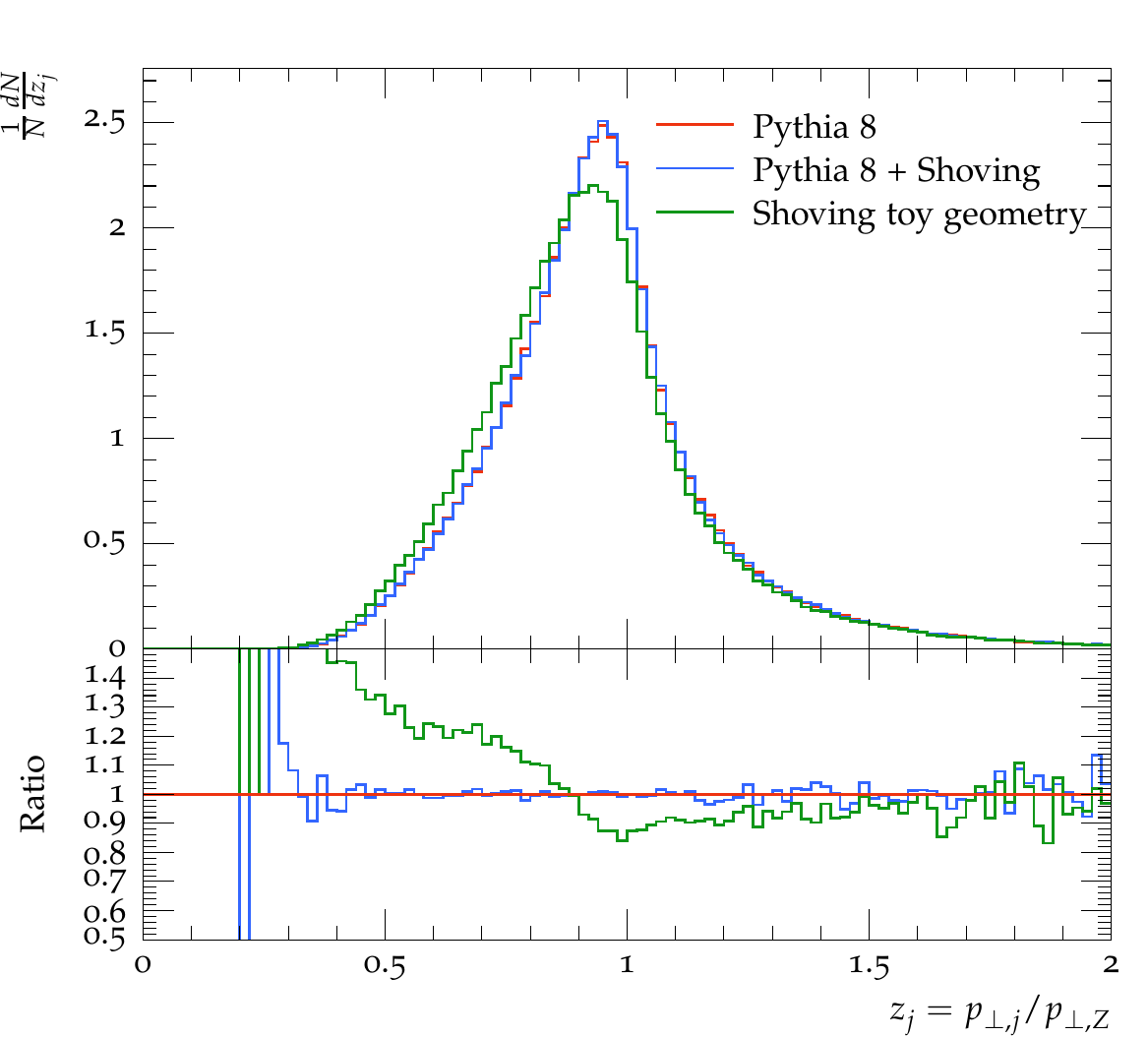}
	\caption{\label{fig:pp-sketch} (a) A sketch showing a high multiplicity pp collision in impact parameter space ($\vec{b}$) and rapidity ($y$), with several MPIs populating the collision with strings. The collision also features a $Z$ boson and a jet. In a normal configuration (black), the hard part of the jet fragments outside the densely populated region. In the used toy geometry (red), the jet is forced to fragment inside the densely populated region. (b) The ratio $z_j = p_{\perp,j}/p_{\perp,z}$ with default \pythia~(red), \pythia~+ shoving with normal event geometry (blue), and the toy event geometry (green).} 
\end{figure*}

\section{The microscopic model for collectivity}
Most general features of pp collisions, such as particle multiplicities and jets, can 
be described by models based on string fragmentation \cite{Andersson:1983jt,Andersson:1979ij}.
In the original model, such strings have no transverse extension, and hadronize
independently. The longitudinal kinematics of the $i$'th breaking is given by the Lund symmetric fragmentation
function:
\begin{equation}
	\label{eq:lufrac}
	f(z) = Nz^{-1}(1-z)^a\exp\left(\frac{-bm_\perp}{z}\right),
\end{equation}
where $z$ is the fraction of the \emph{remaining} available momentum taken away by the hadron. $N$ is a normalization constant, and $a$ and $b$ are tunable parameters, relating the 
fragmentation kinematics to the breakup space-time points of the string, which are located around a hyperbola 
with a proper time of:
\begin{equation}
	\label{eq:stime}
	\langle \tau^2 \rangle = \frac{1+a}{b\kappa^2},
\end{equation}
where $\kappa \sim 1$ GeV/fm is the string tension. The transverse dynamics is determined by the Schwinger result:
\begin{equation}
	\label{eq:schwinger}
	\frac{d\mathcal{P}}{d^2p_\perp} \propto \kappa \exp\left(\frac{\pi m_\perp^2}{\kappa}\right),
\end{equation}
where $m_\perp$ is the transverse mass of the \emph{quark} or \emph{diquark} produced in the string breaking\footnote{The formalism does not dictate whether to use current or constituent quark masses. In \pythia~the suppression factors s/u and diquark/quark are therefore determined from data, with resulting quark masses providing a consistency check.}.

When a $q\bar{q}$ pair moves apart, spanning a string between them, the string length is zero at time $\tau = 0$. To obey causality, its transverse size must also be zero, allowing no interactions between strings for the first short time ($<1$ fm/c) after the initial interaction. After this initial transverse expansion,
strings may interact with each other, by exerting small transverse shoves on each other. In refs.~\cite{Bierlich:2017vhg,Bierlich:2016vgw}
a model for this interaction was outlined, based on early considerations by Abramowski \etal~\cite{Abramovsky:1988zh}. Assuming that 
the energy in a string is dominated by a longitudinal colour--electric field, the transverse interaction force, 
per unit string length is, for two parallel strings, given by:
\begin{equation}
	\label{eq:shoving}
	f(d_\perp) = \frac{g\kappa d_\perp}{\rho^2}\exp\left(-\frac{d_\perp^2}{4\rho^2}\right),
\end{equation}
where both $d_\perp$ (the transverse separation of the two strings), and $\rho$ (the string transverse width) are 
time dependent quantities. The parameter $g$ is a free parameter, which should not deviate too far from unity. 
Equation (\ref{eq:stime}) gives an (average) upper limit for how long time the strings should be allowed to shove each 
other around, as the strings will eventually hadronize\footnote{Eq. (\ref{eq:stime}) is written up with a string
in vacuum in mind. It might be possible that the string life time is modified in the dense environment of a heavy
ion collision.}. String hadronization and the shoving model has been implemented in the \pythia~event generator, and all
predictions in the following are generated using this implementation.

\section{Effects on jet hadronization}
\label{sec:jet-had}
We consider now a reasonably hard $Z$-boson, produced back--to--back with a jet. Due to the large $p_\perp$ 
of the jet, its core will have escaped the transverse region in which shoving takes place well before it is 
affected. See figure \ref{fig:pp-sketch} (a, left) in black for a sketch. In the following, a toy geometry
where the jet is prevented from escaping before shoving, will also be studied, see figure \ref{fig:pp-sketch} 
(a, right) in red for a sketch. The toy geometry is motivated by studies of jet fragmentation in Pb--Pb 
collisions, where the jet must still traverse through a densely populated region before hadronizing, due to
the much larger geometry. Indeed in Pb--Pb, the observed effect by CMS \cite{Sirunyan:2017jic}, is that
the jet-$p_\perp$ relative to the $Z$-$p_\perp$ is reduced, moving the $z_j = p_{\perp, j}/p_{\perp, Z}$ distribution
to the left.

Both geometries are constructed by picking the transverse position of each MPI according to the convolution of the two
proton mass distributions, which are assumed to be 2D Gaussians. The jet is placed in origo. In the first, more realistic,
geometry, all string pieces -- including that corresponding to the jet core -- are allowed to propagate for a finite time,
indicating the time it takes for the strings to from infinitesimal transverse size, to their equilibrium size. In the
toy geometry no such propagation is allowing, and all strings are treated \emph{as if} expanded to full transverse size
at $\tau = 0$. As such, strings from the underlying event, are allowing to shove even the hardest fragment of the jet.
This clearly violates causality, and is not meant to be a realistic picture of a pp interaction. It is
implemented in order to give an effect similar to what one should expect from a heavy-ion collision, where the event geometry 
allows strings from other nucleon-nucleon sub-collisions to interact with the jet core.

A set-up similar to that of the CMS study \cite{Sirunyan:2017jic},
just for pp collisions at $\sqrt{s} = 7$ TeV, is studied in the following. A $Z$-boson reconstructed from leptons with 
$80$ GeV$ < M_Z< 100$ GeV, $p_\perp > 40$ GeV is required, and the leptons are required each to have $p_\perp > 10$ GeV. 
The leading anti-$k_\perp$ \cite{Cacciari:2008gp} jet (using FastJet \cite{Cacciari:2011ma} in 
Rivet \cite{Buckley:2010ar}) is required to have $p_\perp > 80$ GeV and $\Delta \phi_{zj} > 3\pi/4$.
We study three different situations, with the result given in figure \ref{fig:pp-sketch} (b). 

In red, default \pythia~is shown, where geometry has no impact on the result. In blue \pythia~+ shoving, with the normal event 
geometry, with the jet escaping. In green \pythia~+ shoving with the toy geometry, where the jet 
core interacts with the underlying event.

\begin{figure}
\includegraphics[width=0.45\textwidth]{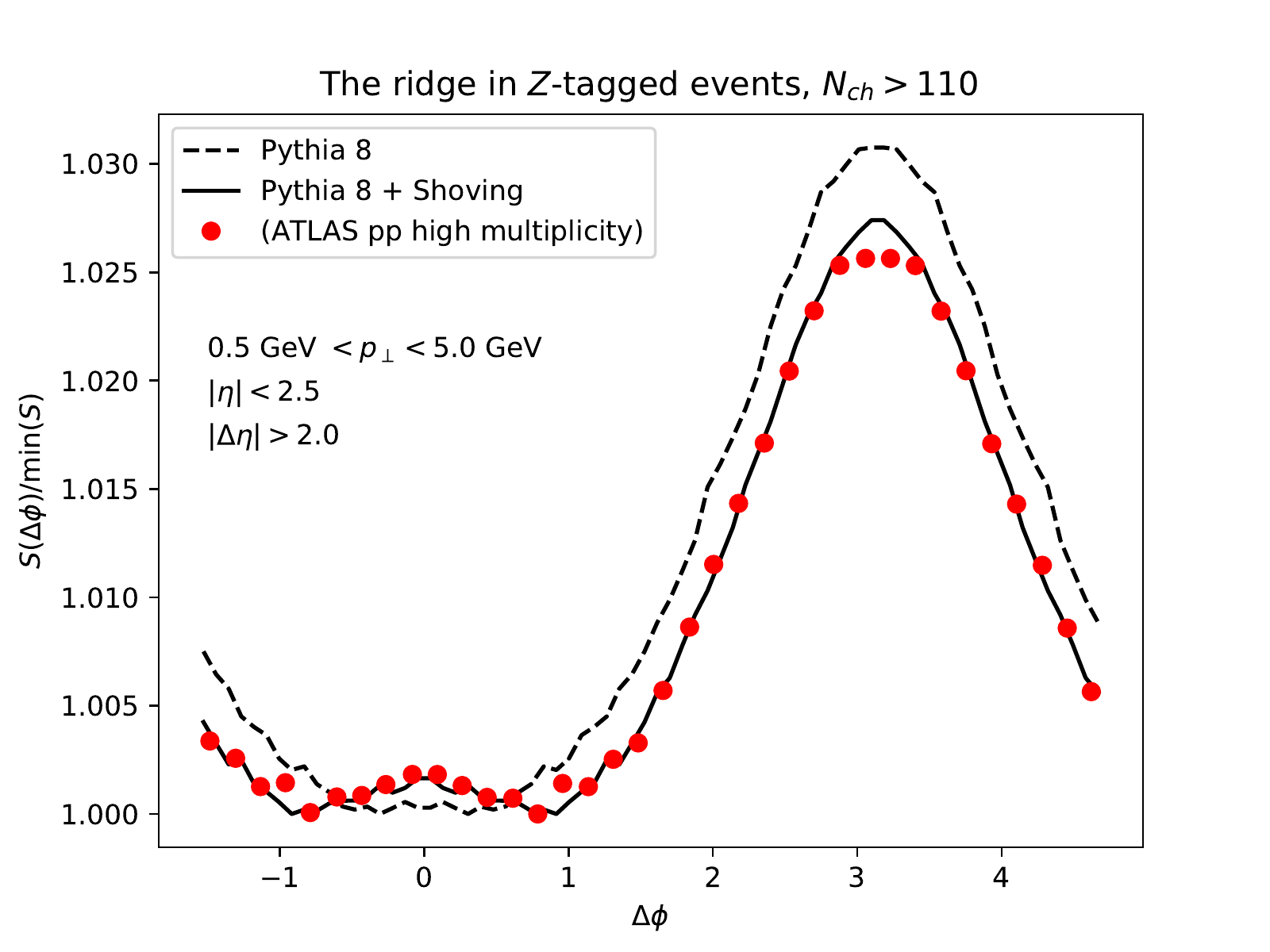}
	\caption{\label{fig:zridge}The ridge in $Z$-tagged, high multiplicity pp collisions at 8 TeV, 
	with default \pythia~(dashed line, no ridge), and \pythia~+ shoving (full line, ridge). 
	Simulation is compared to preliminary ATLAS data \cite{ATLAS:2017nkt}.}
\end{figure}

While shoving in a toy geometry (green) produces an effect qualitatively similar to what one would 
expect from jet quenching, the effect in real events (blue) is far too geometrically suppressed to be seen (comparing blue to red in 
figure \ref{fig:pp-sketch} (b)).
Several suggestions exist for accommodating this problem, prominently using jet substructure observables \cite{Andrews:2018jcm},
or \eg~using a delayed signal from top decays \cite{Apolinario:2017sob} (in AA collisions). In the remaining paper another 
approach will be described. Instead of looking for deviations in the spectrum of a narrow jet compared to
a "vacuum" expectation, we start from the wide-$R$ ($R^2 = \Delta \eta^2 + \Delta \phi^2$) part where collectivity 
in the form of a ridge is known to exist even in pp collisions. The same observable is then calculated as function of
$R$, all the way to the core, where the soft modification is expected to vanish.

\section{Near side ridge in $Z$-tagged events}
The ridge, as recently measured by ATLAS in events with a $Z$ boson present \cite{ATLAS:2017nkt}, provides an opportunity.
The requirement of a $Z$ boson makes the events in question very similar 
to the events studied above. The $Z$ does not influence the effect of the shoving model, and in figure \ref{fig:zridge} 
high multiplicity events, with and without shoving, are shown, with the appearance of a ridge in the latter -- in accordance 
with the experimental results\footnote{The simulation is compared to preliminary ATLAS data \cite{ATLAS:2017nkt}, with the caveat 
that the analysis procedure is very simplistic compared to the experimental one. Instead of mixing signal events with a background sample, distributions are instead divided each with their minimum to obtain comparable scales.}.

It is instructive to discuss the result of figure \ref{fig:zridge} with the sketch in figure \ref{fig:pp-sketch} (a)
in mind. Since the ridge analysis requires a $|\Delta \eta|$ gap of 2.5, the jet region is, by construction, cut away.
(Keeping in mind that in this case there is no required jet trigger.)
The underlying event does, however, continue through the central rapidity range, and if only one could perform a true separation
of jet particles from the underlying event in an experiment, the ridge should be visible. Since that is not possible, it is reasonable to 
naively ask if the presence of a ridge in the underlying event will by itself give rise to a shift in $z_j$. The result presented in figure 
\ref{fig:pp-sketch} (b) (blue line) suggests that it does not. It is therefore necessary to explore more exclusive observables to isolate
the effect of the soft modification of the jet.

\begin{figure}
	\includegraphics[width=0.45\textwidth]{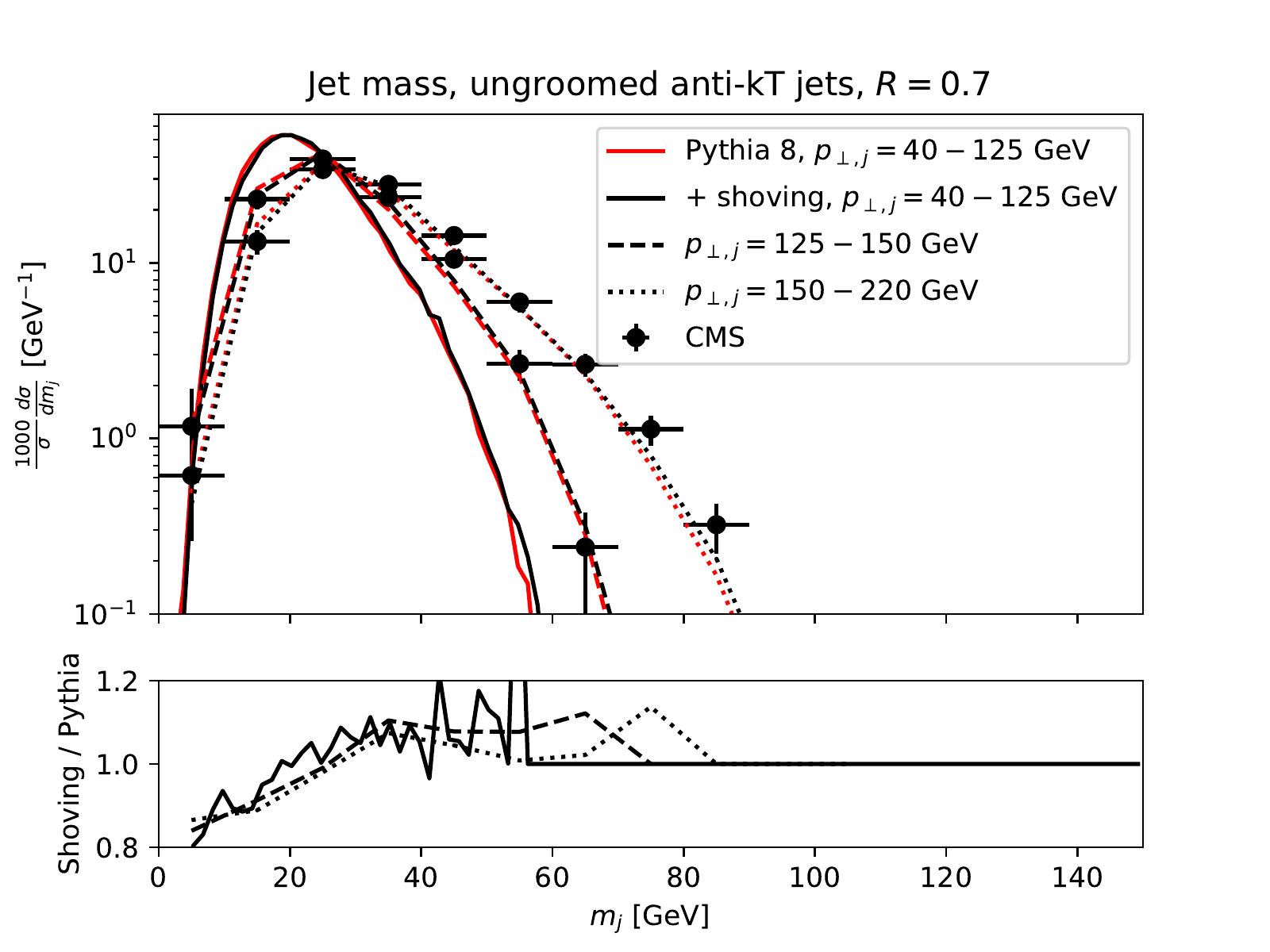}
	\caption{\label{fig:jetmass} The jet mass of anti-$k_\perp$ jets with $R = 0.7$, in events with a $Z-$boson with $p_\perp > 120$ GeV, in bins of jet-$p_\perp$. Data compared to default \pythia~(red) and \pythia~+ shoving (black).} 
\end{figure}

The comparison in figure \ref{fig:zridge} also serves the purpose of fixing the parameters of the shoving model before studying jet-related 
quantities. The only free parameter of the model is the $g$-parameter in equation (\ref{eq:shoving}), the rest are fixed to default values \cite{Skands:2014pea}. As shown in ref. \cite{Bierlich:2017vhg}, the free parameter determines the height of the ridge. The value $g=4$ is chosen in this paper, which also gives a good description of the ridge in minimum bias events.

\section{Influence on jet observables}
As the ATLAS measurement has established, there is indeed collectivity present in (high multiplicity) events with a $Z$ present. 
In the previous section it was shown that the measured signature can be adequately 
described by the shoving model. Now the situation will be extended to include also a high-$p_\perp$ jet trigger in the
same way as in section \ref{sec:jet-had}, and the effect of the collective behaviour on the jet will be discussed.

\subsection{Hard measures: Jet mass and jet cross section}
The jet masses, binned in jet-$p_\perp$, is a key calorimetric observable for comparing observed jet
properties to predictions from models. With the advent of jet grooming techniques, the 
precision of such comparisons have increased, and any model seeking to predict new phenomena, must
be required to not destroy any previous agreement with this observable. The mass of hard 
jets produced in events with a $Z$-boson present has been measured by the CMS experiment 
\cite{Chatrchyan:2013vbb}, and in figure \ref{fig:jetmass} the results are compared to \pythia with and without shoving, in red and black respectively.
Shoving increases the jet mass slightly, bringing the prediction closer to data, though not at a significant level.
In the analysis by CMS \cite{Chatrchyan:2013vbb}, various 
grooming techniques are also explored. These are not shown in the figure, but all remove most of the effect from shoving.
This is the expected result, as the grooming techniques are in fact introduced to remove soft QCD radiation from jets.

\begin{figure}
\includegraphics[width=0.45\textwidth]{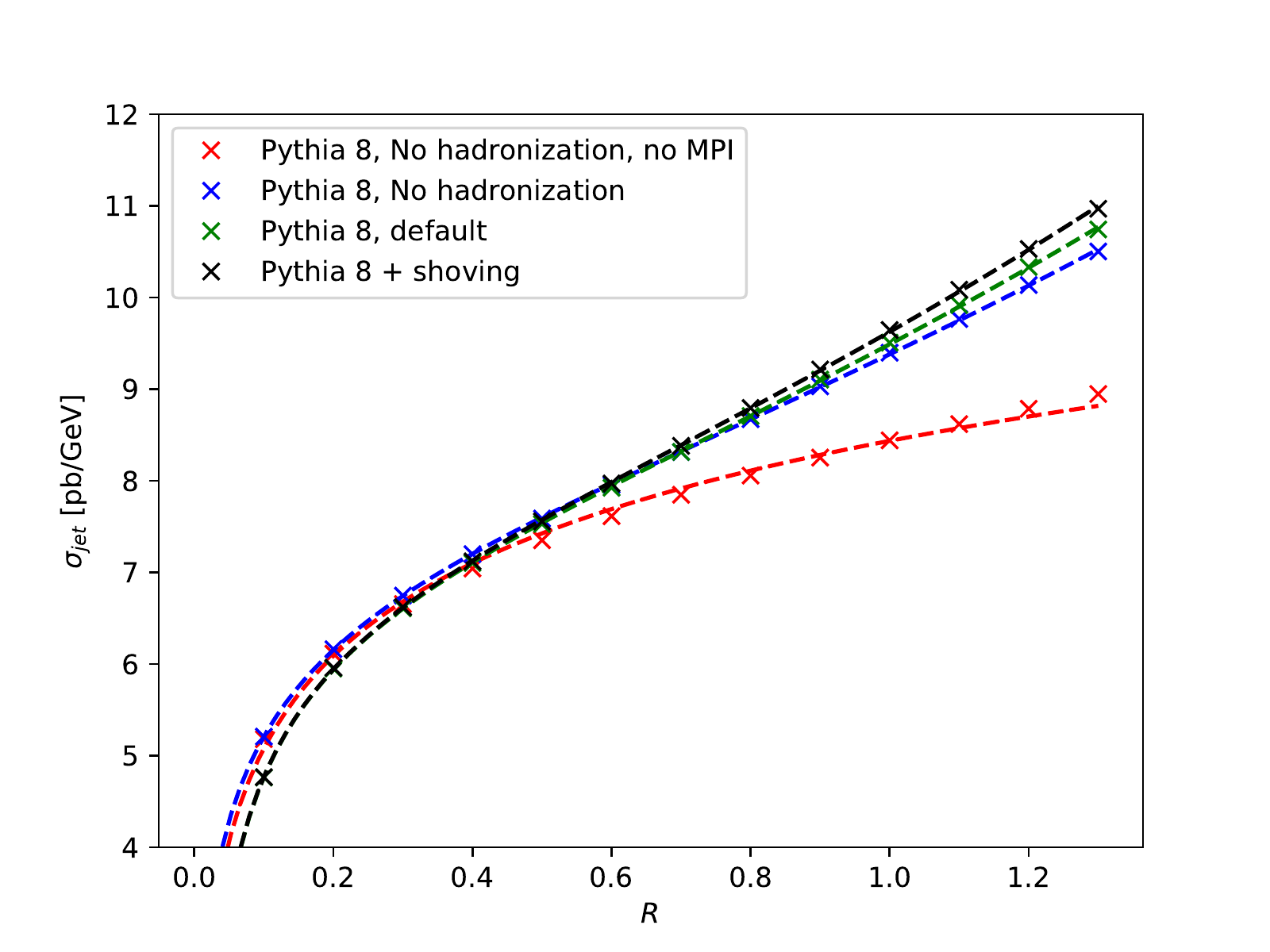}
	\caption{\label{fig:jetr}The $R$ dependence of $\sigma_j$ for four configurations of the leading 
	jet in $Z+$jet in pp collisions at 7 TeV. Special attention is given to the difference between \pythia~ 
	default and \pythia~+ shoving in the large-$R$ limit.}
\end{figure}

An effect of shoving at the \mbox{10 \%} level is seen for low jet masses. While also the most difficult region 
to assess experimentally, this effect could be worthwhile to explore further. A prediction for the jet-$p_\perp$ bin 40-125 GeV is also shown, as one could imagine that a larger effect could be observed if the jet threshold could be experimentally lowered. The effect on a \mbox{10 \%} level persists, but does not increase.

The jet-$p_\perp$ is also a well studied quantity. As there is little effect on the raw jet-$p_\perp$ spectra, the jet cross section is used:
\begin{equation}
	\sigma_j = \int_{p_{\perp,0}}^\infty dp_{\perp,j} \frac{d\sigma}{dp_{\perp,j}},
\end{equation}
where $p_{\perp,j}$ is the $p_\perp$ of the leading jet in the event, and $p_{\perp, 0}$ is the imposed phase space cut--off.
It was pointed out by Ellis~\etal~\cite{Ellis:1992qq}, that the $R$-dependence of $\sigma_j$ under the influence of MPIs in a pp collision, can be parametrized as $A + B\log(R) + CR^2$. 
Later Dasgupta \etal~\cite{Dasgupta:2007wa} noted that hadronization effects contributes like $-1/R$. This gives a total parametrization:
\begin{equation}
	\label{eq:sigmaj}
	\sigma_j(R) = A + B\log(R) + CR^2 - DR^{-1}.
\end{equation}
By construction, the ridge effect from the previous chapter is far away from the jet in $\eta$, and therefore also in $R$. 
Any contribution from shoving can be reasonably expected to be most pronounced for large $R$. Equation (\ref{eq:shoving}) gives a contribution
of $\langle dp_\perp / d\eta \rangle \propto f\left(\langle d_\perp \rangle \right)$, where $\langle d_\perp \rangle$ is
density dependent. In the previously introduced semi-realistic geometry, a contribution to $\sigma_j$, which
is $\propto R^2$, is expected, \ie~a correction to the parameter $C$ in equation (\ref{eq:sigmaj}).

In figure \ref{fig:jetr}, $\sigma_j(R)$ is shown without MPIs and hadronization (red), with MPI, no hadronization (blue), 
\pythia~default (green) and \pythia~+ shoving (black). The analysis setup is the same as in section \ref{sec:jet-had}.
Results from the Monte Carlo is shown as crosses, and the resulting fits as dashed lines, with parameters given in table \ref{partable}.

\begin{table}
{\footnotesize
\begin{center}
\begin{tabular}{lcccc}
   [pb/GeV] & No MPI, no had. & No had. & Default & Shoving \\
    \hline
    A & $1.46 \pm 0.03$ & $1.31 \pm 0.01$ & $1.28 \pm 0.04$ & $1.29 \pm 0.05$ \\
    B & $8.44 \pm 0.03$ & $8.22 \pm 0.01$ & $8.18 \pm 0.02$ & $8.19 \pm 0.03$ \\
    C & - & $1.16 \pm 0.01$ & $1.35 \pm 0.03$ & $1.49 \pm 0.03$ \\
    D & - & - & $0.05 \pm 0.01$ & $0.05 \pm 0.01$ \\
    \hline
\end{tabular}
\end{center}
}
\caption{\label{partable}Parameters obtained by fitting equation (\ref{eq:sigmaj}) to \pythia. Errors are fit errors ($1\sigma$), 
	fits shown in figure \ref{fig:jetr}.}
\end{table}
From the fits it is visible that shoving contributes to the $R^2$ dependence as expected. Directly from figure \ref{fig:jetr} it is
visible that shoving contributes to the jet cross section at a level comparable to hadronization effects. As it is also seen from the
figure and table, MPI effects contributes much more than the additional effects from hadronization or shoving. This means that the
usual type of centrality measure (number of charged particles measured in some fiducial region) is not quite applicable for such observables,
as the large bias imposed on MPI selection, would overcome any bias imposed on selection of the much smaller collective effects\footnote{In order to use this procedure to set limits on jet quenching in small systems, comparison must be made to predictions. 
In figure \ref{fig:jetr} only LO predictions are given, but while NLO corrections are sizeable enough that figure \ref{fig:jetr} 
cannot be taken as a numerically accurate prediction, such corrections will not affect the relative change in $\sigma_j$ from shoving, 
and will not affect the result. More crucial is the effect of parton density uncertainties, which may 
affect $\sigma_j$ up to 10\% for this process \cite{Bendavid:2018nar}. This points to the necessity of more precise determinations of PDFs, 
if microscopic non-perturbative effects on hard probes in pp collisions are to be fully understood.}

\begin{figure}
	\includegraphics[width=0.45\textwidth]{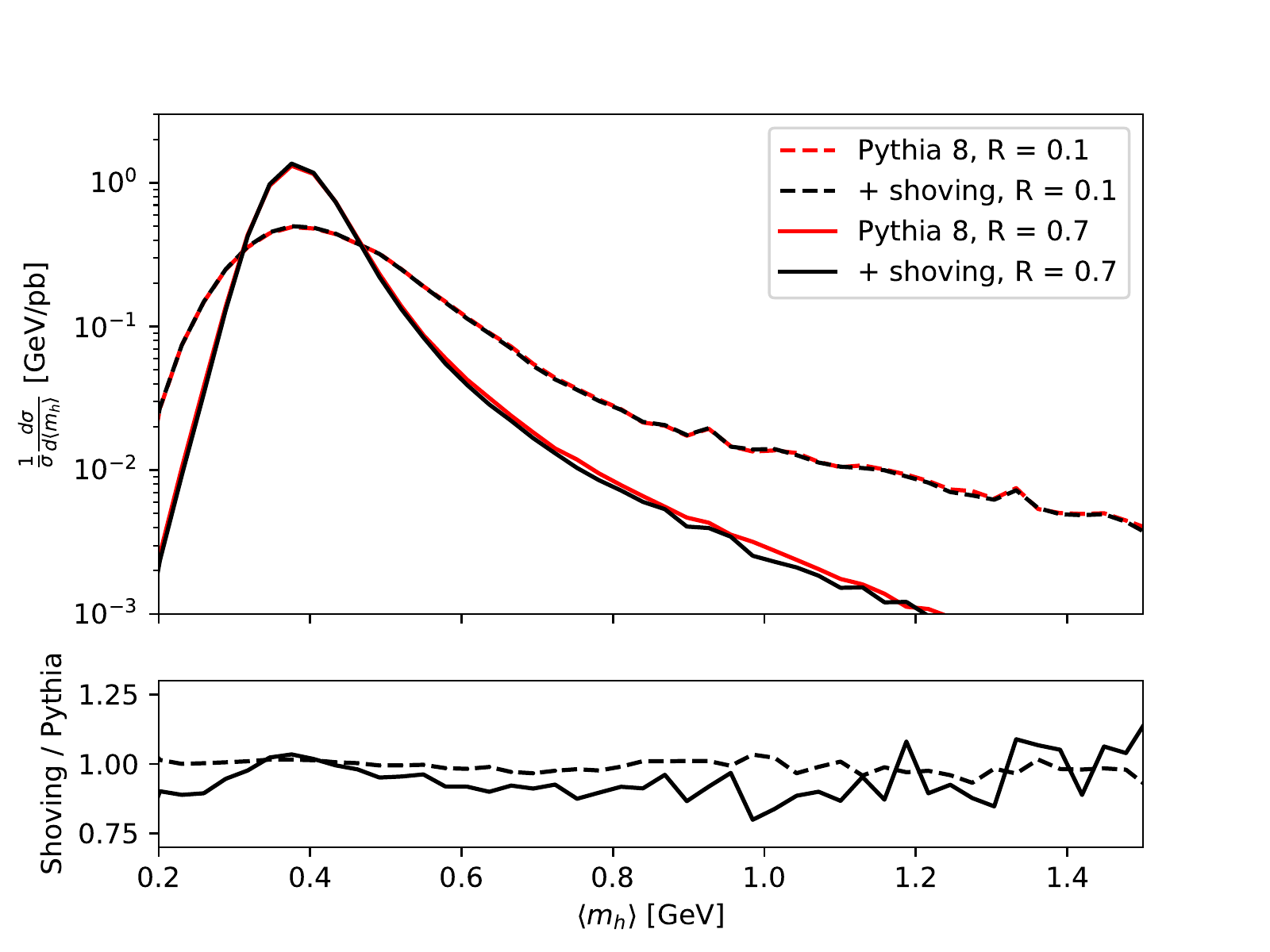}
	\caption{\label{fig:avgmass} The average hadron mass in the leading anti-$k_\perp$ jet with $R = 0.1$ (dashed) and $R = 0.7$ (full) in $Z$+jet, using default \pythia~(red) and \pythia~+ shoving (black). The deviation imposed by shoving grows larger with increasing $R$.} 
\end{figure}

\subsection{Soft measures: Average hadron mass and charge}
The hadrochemistry of the jet is here quantified in a quite inclusive manner by the average hadron mass:
\begin{equation}
	\langle m_h \rangle = \frac{1}{N_p} \sum_i^{N_p} m_{h,i},
\end{equation}
where $N_p$ is the number of hadrons in the jet, and $m_h$ are the individual hadron masses. Furthermore the total jet charge is studied:
\begin{equation}
	Q_j = \sum_i^{N_p} q_{h,i},
\end{equation}
where $q_i$ are the individual hadron electric charges. As shoving only affects these quantities indirectly, the predicted
effect is not as straight forward as was the case for jet cross section, but requires a full simulation to provide predictions.
In figure \ref{fig:avgmass} the average hadron mass in the leading jet (still in $Z$+jet collisions as above) is shown for 
two exemplary values of $R$. For small $R$, $\langle m_h \rangle$ is unchanged, but as $R$ grows, a significant change, on the order 
of \mbox{10 \%} is visible. The average hadron mass in jets has to this authors' knowledge not been measured inclusively, but related quantities (ratios of particle
species) has been preliminarily shown by ALICE \cite{Hess:2014xba} to be adequately described by \pythia.

\begin{figure}
	\includegraphics[width=0.4\textwidth]{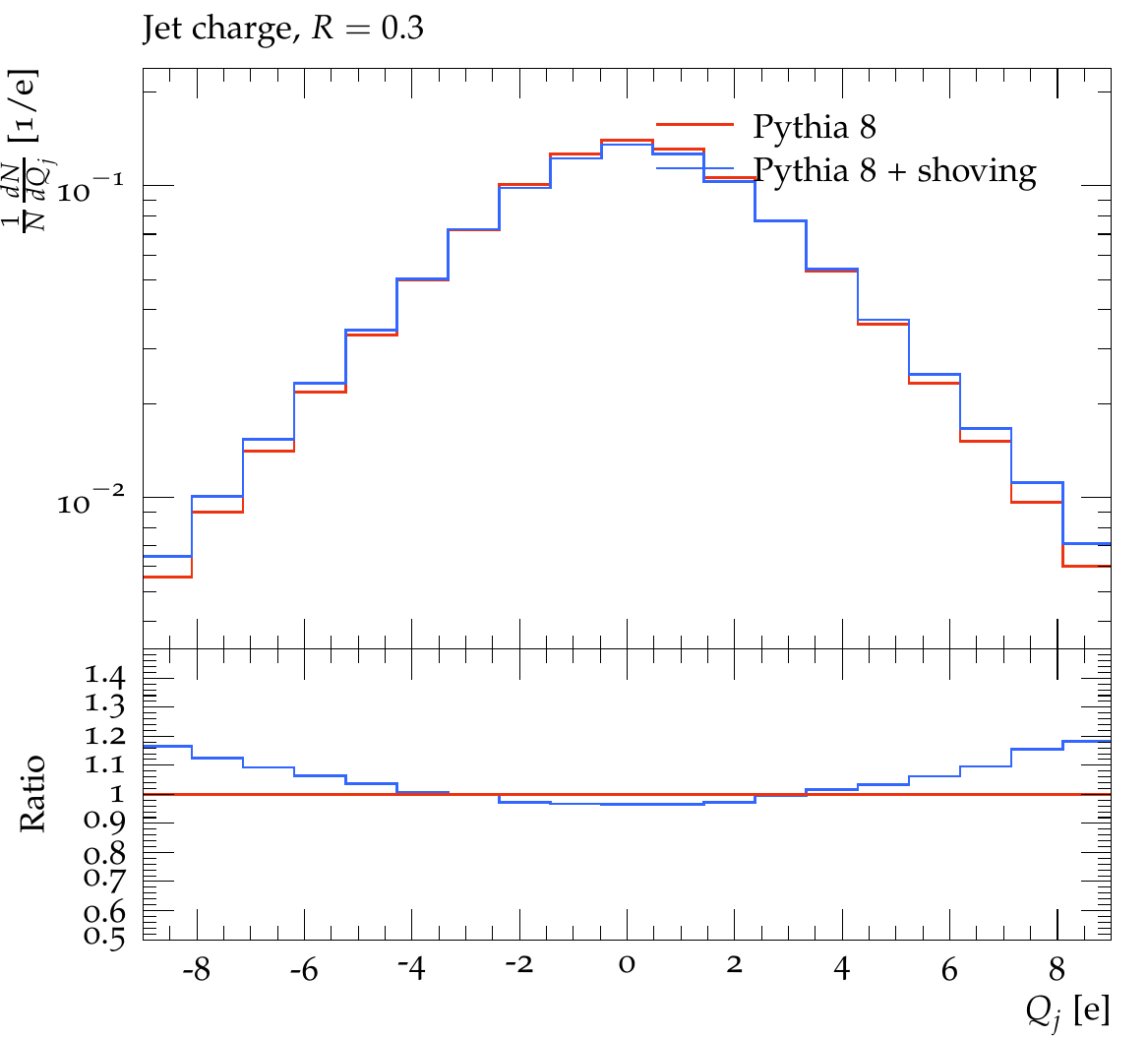}
	\caption{\label{fig:jetchargedist} An example of a jet charge distribution for the leading anti-$k_\perp$ jet in $Z$+jet with $R = 0.3$. Shoving has the effect of making the distribution wider.} 
\end{figure}

\begin{figure}
	\includegraphics[width=0.45\textwidth]{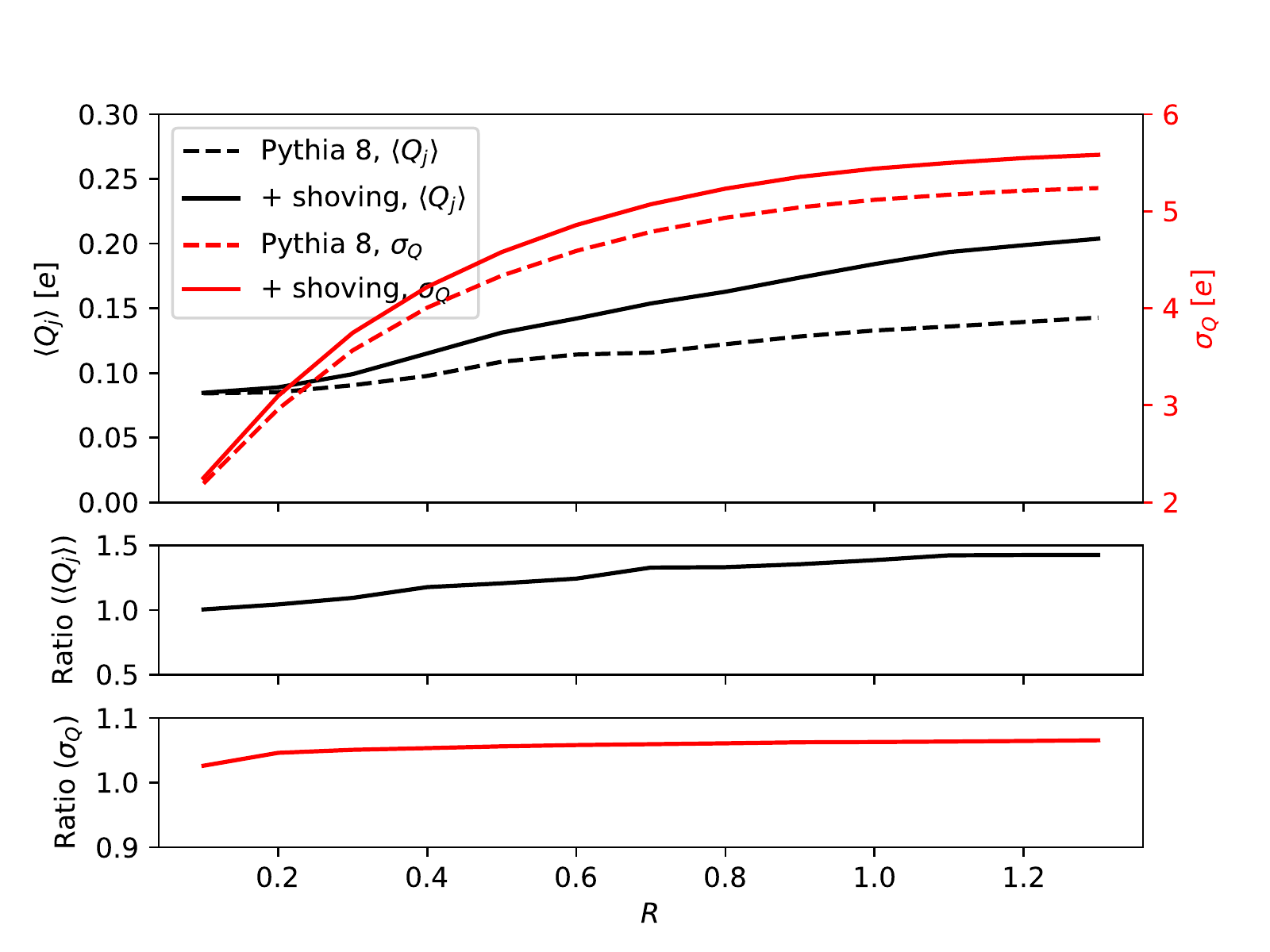}
	\caption{\label{fig:jetcharger} $R$-dependency of the average jet charge and the distribution (see fig. \ref{fig:jetchargedist}) width with and without shoving. Note the different scales for the two quantities.} 
\end{figure}

The $Q_j$ distribution for $R=0.3$ jets is shown in figure \ref{fig:jetchargedist}. It is seen directly, that for this particular value 
of $R$, shoving widens the distribution, and also the mean is further shifted in the positive direction. The $R$-dependency of this behaviour is shown in figure \ref{fig:jetcharger}. Here both the mean and the width of the jet distribution at different values of $R$ is shown (note the different scales on the axes). It is seen that this observables shows deviations up to \mbox{40 \%} in the large-$R$ limit. Jet identification techniques to reveal whether the seed parton is a gluon or a quark \cite{Gras:2017jty,Bright-Thonney:2018mxq} might be able to increase the discriminatory power even further.

\pythia~provides a good description of jet charge in di-jet events \cite{Sirunyan:2017tyr}, giving further significance to any deviation introduced by shoving in this special configuration. It should, however, be noted that the jet charge has been a challenge for fragmentation models since the days of $e^+e^-$ collisions at LEP \cite{Abdallah:2006ve}. The renewed interest in fragmentation properties from the observation of collectivity in small systems provides a good opportunity to also go back and revisit older observations.

The jet hadrochemistry can be studied in a more exclusive manner, by means of particle identification, similar to what is done in nuclear collisions. Such observables will also be largely affected by formation of colour multiplets, increasing the string tension \cite{Biro:1984cf,Bierlich:2014xba}. In the context of this letter, it is noted that rope formation contributes negligibly to the observables studied above. Some studies of rope effect in jets in pp collisions have been performed \cite{Mangano:2017plv}, but could require further attention to the important space--time structure, as described in section \ref{sec:jet-had}. 

\section{Conclusions}
The non-observation of jet quenching in small systems is one of the key open questions to understand collective behaviour in collisions of protons. For the coming high luminosity era at LHC, the search for new observables to either observe jet quenching, or provide quantitative exclusion limits is necessary. In this letter we have shown that the microscopic model for collectivity implemented in \pythia, can reproduce one observed collective feature already observed in pp collisions with a hard probe, namely the ridge in $Z$ tagged events. Basic features like $z_j$ are, however, unaffected, but highly sensitive to the collision geometry. For a toy event geometry, the model produces features similar to those observed in Pb--Pb collisions. The toy geometry study highlights the need for a better motivated theoretical description of the space--time structure of the initial state. The realization that the complicated interplay between fragmentation time and spatial structure is significant for precision predictions, dates back to the 1980's for collisions of nuclei \cite{Bialas:1986cf}. With the discovery of small system collectivity, several approaches have been developed also for pp collisions (\textit{e.g.} \cite{Avsar:2010rf,dEnterria:2010xip,Albacete:2016pmp,Ferreres-Sole:2018vgo}), most (but not all) aiming for a description of flow effects. It is crucial for the future efforts that such space--time models attempt at describing both soft and hard observables at once, in order to avoid "over tuning" of sensitive parameters. In this letter it was done by first describing the ridge in $Z$-tagged events, and then proceed to investigate jet observables with the same parameters, while the models remains able to describe key observables like jet mass. An effect from shoving up to \mbox{10 \%} for low jet masses was shown, but is within the current experimental uncertainty.

The major contribution of this letter is the proposal of several new observables to understand the effects on jet fragmentation from the shoving model in $Z$+jet events. The main idea behind these observables is to go from the wide-$R$ region (wide jets), where collective effects, in form of the ridge, is already observed, to the very core of the jet, where only little effect is expected. The jet-$p_\perp$ is only affected little, and the observed \mbox{5 \%} effect on the integrated quantity $\sigma_j$, will be difficult to observe when also taking into account uncertainties from PDFs and NLO corrections, but nevertheless provides a crucial challenge for the upcoming high luminosity experiments at LHC, where larger statistics can help constraining the theoretical uncertainties better. 
More promising are the effects observed on hadron properties inside the jet, where the average hadron mass shows a \mbox{10 \%} deviation and jet charge even larger. Even if an effect this large is not observed in experiment, its non-observation will aide the understanding of soft collective effects better, as the shoving model predicting the effect, adequately describes the ridge in $Z$-tagged collisions.

\section{Acknowledgements}
I thank Johannes Bellm for valuable discussions, and Peter Christiansen, Leif L\"{o}nnblad and G\"{o}sta Gustafson for critical comments on the manuscript. I am grateful for the hospitality extended to me by the ALICE group at the Niels Bohr Institute during the preparation of this work.
This work was funded in part by the Swedish Research Council, contract number 2017-0034, and in part by the MCnetITN3 H2020 Marie Curie Initial Training Network, contract 722104.

\bibliographystyle{elsarticle-num}
\bibliography{jetq}

\end{document}